\documentclass[aps,prd,onecolumn,groupedaddress,showpacs,nofootinbib,amssymb]{revtex4}
\usepackage[dvips]{graphicx}
\usepackage{amssymb}
\usepackage{amsmath}
\usepackage{graphicx,,color}
\usepackage{amsfonts}
\usepackage{bm}
\usepackage{cancel}
\usepackage{comment}

\newcommand\be{\begin{equation}}
\newcommand\ee{\end{equation}}

\allowdisplaybreaks[4]

\begin{document}

\tolerance=5000

\title{Bottom-Up Reconstruction of Viable GW170817 Compatible Einstein-Gauss-Bonnet Theories}
\author{V.K.~Oikonomou,$^{1,2}$}
\email{v.k.oikonomou1979@gmail.com,voikonomou@auth.gr}
\author{P.D. Katzanis,$^{1}$}
\email{pkatzanis@gmail.com,pkatzani@auth.gr}
\author{Ilias C. Papadimitriou,$^{1}$}
\email{elias.papajim@gmail.com,ipapadim@auth.gr}
\affiliation{$^{1)}$ Department of Physics, Aristotle University of Thessaloniki, Thessaloniki 54124, Greece\\
$^{2)}$ Laboratory for Theoretical Cosmology, International Center
of Gravity and Cosmos, Tomsk State University of Control Systems
and Radioelectronics, 634050 Tomsk, Russia (TUSUR)}

\tolerance=5000

\begin{abstract}
In this work we shall use a bottom-up approach for obtaining
viable inflationary Einstein-Gauss-Bonnet models which are also
compatible with the GW170817 event. Specifically, we shall use a
recently developed theoretical framework in which we shall specify
only the tensor-to-scalar ratio, in terms of the $e$-foldings
number. Starting from the tensor-to-scalar ratio, we shall
reconstruct from it the Einstein-Gauss-Bonnet theory which can
yield such a tensor-to-scalar ratio, finding the scalar potential
and the Gauss-Bonnet coupling scalar function as functions of the
$e$-foldings number. Accordingly, the calculation of the spectral
index of the primordial scalar perturbations, and of the tensor
spectral index easily is greatly simplified and these
observational indices can easily be found. After presenting the
general formalism for the bottom-up reconstruction, we exemplify
our findings by presenting several Einstein-Gauss-Bonnet models of
interest which yield a viable inflationary phenomenology. These
models have also an interesting common characteristic, which is a
blue tilted tensor spectral index. We also investigate the
predicted energy spectrum of the primordial gravitational waves
for these Einstein-Gauss-Bonnet models, and as we show, all the
models yield a detectable primordial wave energy power spectrum.
\end{abstract}


\maketitle

\section{Introduction}

In the next two decades, the cosmologists's community anticipates
a great amount of information coming from the sky. This
information will either validate the current state of art thinking
on cosmological issues, or will exclude current theories and thus
our perception about the early Universe will utterly change.
Indeed, in about 10-15 years from now, several experiments that
are based on astronomical observations will commence to yield
their first data, like for example LISA
\cite{Baker:2019nia,Smith:2019wny}. But also several other future
proposed experiments might also begin to operate, like BBO
\cite{Crowder:2005nr,Smith:2016jqs} and DECIGO
\cite{Seto:2001qf,Kawamura:2020pcg}. During the 2010's decade, the
cosmological community has already experienced the first surprises
coming from the sky. Specifically, the GW170817 kilonova
accompanied event
\cite{TheLIGOScientific:2017qsa,Monitor:2017mdv,GBM:2017lvd},
imposed severe constraints to theories that predict tensor
spacetime perturbations with gravitational wave speed $c_T^2$
different from unity in natural units. Specifically, the
electromagnetic counterpart to GW170817 indicates that the
deviation in the speed of gravitational waves from that of light
must roughly be $|c_T/c-1|\leq 5 \times 10^{-16}$, see for example
Refs.
\cite{Ezquiaga:2017ekz,Baker:2017hug,Creminelli:2017sry,Sakstein:2017xjx,Fernandes:2022zrq}.
One relevant example in which case the theory predicts $c_T^2 \neq
1$, while comfortably well fitted within observational
constraints, are the scalar-tensor formulations of
four-dimensional Einstein-Gauss-Bonnet gravity, see for example
the recent review \cite{Fernandes:2022zrq}. In these models, the
GW170817 event imposes only mild constraints on the coupling
constant of the theory. For more details on the constraints
imposed by the GW170817 event on several versions of modified
gravity and scalar tensor gravity, see for example
\cite{Ezquiaga:2017ekz,Baker:2017hug,Creminelli:2017sry,Sakstein:2017xjx,Fernandes:2022zrq}.
In recent works \cite{Odintsov:2020sqy,Oikonomou:2021kql} we
provided a theoretical framework for Einstein-Gauss-Bonnet
theories
\cite{Hwang:2005hb,Nojiri:2006je,Cognola:2006sp,Nojiri:2005vv,Nojiri:2005jg,Satoh:2007gn,Bamba:2014zoa,Yi:2018gse,Guo:2009uk,Guo:2010jr,Jiang:2013gza,Kanti:2015pda,vandeBruck:2017voa,Kanti:1998jd,Dominguez:2005rt,Maeda:2005ci,Ghosh:2011ad,Maharaj:2015gsd,Brassel:2019bam,Pozdeeva:2020apf,Vernov:2021hxo,Pozdeeva:2021iwc,Koh:2014bka,Bayarsaikhan:2020jww,Tumurtushaa:2018lnv,Fomin:2020hfh,DeLaurentis:2015fea,Chervon:2019sey,Nozari:2017rta,Odintsov:2018zhw,Kawai:1998ab,Yi:2018dhl,vandeBruck:2016xvt,Kleihaus:2019rbg,Bakopoulos:2019tvc,Maeda:2011zn,Bakopoulos:2020dfg,Ai:2020peo,Oikonomou:2020oil,Odintsov:2020xji,Oikonomou:2020sij,Odintsov:2020zkl,Odintsov:2020mkz,Venikoudis:2021irr,Kong:2021qiu,Easther:1996yd,Antoniadis:1993jc,Antoniadis:1990uu,Kanti:1995vq,Kanti:1997br}
which yields GW170817-compatible models with gravitational wave
speed $c_T^2\simeq 1$ in natural units, thus equal to that of
light's. Einstein-Gauss-Bonnet theories are interesting on their
own, since these are generalizations of the Einstein-Hilbert
action, which contains linear forms of curvature. In contrast,
Einstein-Gauss-Bonnet theories contain more than just linear forms
of the Riemann tensor, the Ricci tensor and the Ricci scalar.
Basically, Einstein-Gauss-Bonnet gravity is a specific case of
Lovelock gravity \cite{Lovelock:1971yv,Lovelock:1972vz}. Lovelock
gravity applies to higher order and higher dimensional theories of
gravity, and Einstein-Gauss-Bonnet gravity is the second-order
limiting case of general Lovelock's theory while Einstein's
general relativity, is the first-order Lovelock theory. It is a
well known fact that Einstein-Gauss-Bonnet gravity has a minimum
dimension of five, whereas general relativity has a minimum
dimension of four. So effectively Einstein-Gauss-Bonnet gravity
has a general relativity limit but only in five dimensions.

In this work we aim to provide a bottom-up reconstruction
technique for viable Einstein-Gauss-Bonnet theories developed in
\cite{Oikonomou:2021kql}. We shall use the theoretical framework
developed in \cite{Oikonomou:2021kql}, and starting from the
tensor-to-scalar ratio given as a desired function of the
$e$-foldings number, we shall reconstruct the
Einstein-Gauss-Bonnet theory which can yield such a
tensor-to-scalar ratio. Specifically, we shall find the scalar
potential and the Gauss-Bonnet scalar coupling function, as
functions of the $e$-foldings number. From these, the calculation
of the scalar and tensor spectral indices easily follows, and thus
the whole framework is available for checking the viability of
several Einstein-Gauss-Bonnet models. We shall exemplify the
techniques of the bottom-up reconstruction formalism, by
presenting several viable inflationary Einstein-Gauss-Bonnet
models. Also, due to the fact that these models result to a
positive tensor spectral index, we shall also calculate the
predicted primordial gravitational wave energy spectrum for each
of these models. As we show, these models can yield a detectable
signal of primordial gravitational waves. Another interesting
feature of the models which we shall present is the fact that some
of the models can yield a small tensor-to-scalar ratio, which can
be useful in the future, if the stage 4 Cosmic Microwave
Background experiments further constraint the tensor-to-scalar
ratio to smaller values.

This paper is organized as follows: In section II we overview in
brief the theoretical framework of GW170817-compatible
Einstein-Gauss-Bonnet theories developed in
\cite{Oikonomou:2021kql}. In section III we present our bottom-up
reconstruction approach in order to study phenomenologically
viable Einstein-Gauss-Bonnet theories. In section IV we present
several illustrative examples of viable Einstein-Gauss-Bonnet
models using our bottom up reconstruction techniques, and we also
present the potential ability of these model to generate a
detectable energy spectrum of primordial gravitational waves.
Finally, the conclusions follow at the end of the paper.

\section{Brief Overview of GW170817-Compatible Einstein-Gauss-Bonnet Theories}

In this section we shall briefly review the reformed
GW170817-compatible formalism for Einstein-Gauss-Bonnet theories
developed in Ref. \cite{Oikonomou:2021kql}. The vacuum
Einstein-Gauss-Bonnet gravity action has the following form,
\begin{equation}
\label{action} \centering
S=\int{d^4x\sqrt{-g}\left(\frac{R}{2\kappa^2}-\frac{1}{2}\partial_{\mu}\phi\partial^{\mu}\phi-V(\phi)-\frac{1}{2}\xi(\phi)\mathcal{G}\right)}\,
,
\end{equation}
with $R$ denoting the Ricci scalar, $\kappa=\frac{1}{M_p}$ with
$M_p$ being the reduced Planck mass. Moreover, $\mathcal{G}$
denotes the Gauss-Bonnet invariant in dimension-4, which is
$\mathcal{G}=R^2-4R_{\alpha\beta}R^{\alpha\beta}+R_{\alpha\beta\gamma\delta}R^{\alpha\beta\gamma\delta}$
where $R_{\alpha\beta}$ and $R_{\alpha\beta\gamma\delta}$ denote
the Ricci and Riemann tensor respectively. Also for the rest of
this paper we shall assume that the geometry of spacetime is
described by a flat Friedmann-Robertson-Walker (FRW) metric, with
line element,
\begin{equation}
\label{metric} \centering
ds^2=-dt^2+a(t)^2\sum_{i=1}^{3}{(dx^{i})^2}\, ,
\end{equation}
where $a$ denotes the scale factor and also for the FRW metric,
the Gauss-Bonnet invariant takes the form $\mathcal{G}=24H^2(\dot
H+H^2)$. Furthermore, we shall assume that the scalar field is
solely time-dependent. The field equations are easily derived by
varying the gravitational action with respect to the metric and to
the scalar field, and these are,
\begin{equation}
\label{motion1} \centering
\frac{3H^2}{\kappa^2}=\frac{1}{2}\dot\phi^2+V+12 \dot\xi H^3\, ,
\end{equation}
\begin{equation}
\label{motion2} \centering \frac{2\dot
H}{\kappa^2}=-\dot\phi^2+4\ddot\xi H^2+8\dot\xi H\dot H-4\dot\xi
H^3\, ,
\end{equation}
\begin{equation}
\label{motion3} \centering \ddot\phi+3H\dot\phi+V'+12 \xi'H^2(\dot
H+H^2)=0\, ,
\end{equation}
where the ``dot'' denotes differentiations with respect to the
cosmic time $t$. Following Ref. \cite{Oikonomou:2021kql}, by
imposing the slow-roll conditions,
\begin{equation}\label{slowrollhubble}
\dot{H}\ll H^2,\,\,\ \frac{\dot\phi^2}{2} \ll V,\,\,\,\ddot\phi\ll
3 H\dot\phi\, .
\end{equation}
and also the constraint $c_T^2=1$, where $c_T^2$ is,
\begin{equation}
\label{GW} \centering c_T^2=1-\frac{Q_f}{2Q_t}\, ,
\end{equation}
and $Q_f$, $Q_b$ appearing above are $Q_f=8 (\ddot\xi-H\dot\xi)$,
$Q_t=F+\frac{Q_b}{2}$, where  $F$ is equal to
$F=\frac{1}{\kappa^2}$ while $Q_b=-8 \dot\xi H$, the field
equations are simplified as follows,
\begin{equation}
\label{motion5} \centering H^2\simeq\frac{\kappa^2V}{3}\, ,
\end{equation}
\begin{equation}
\label{motion6} \centering \dot H\simeq-\frac{1}{2}\kappa^2
\dot\phi^2\, ,
\end{equation}
\begin{equation}
\label{motion8} \centering \dot\phi\simeq\frac{H\xi'}{\xi''}\, .
\end{equation}
Moreover, the scalar potential and the Gauss-Bonnet scalar
coupling function must satisfy the following constraint
differential equation,
\begin{equation}
\label{maindiffeqnnew} \centering
\frac{V'}{V^2}+\frac{4\kappa^4}{3}\xi'\simeq 0\, .
\end{equation}
The slow-roll indices for Einstein-Gauss-Bonnet models are
\cite{Hwang:2005hb},
\begin{align}\label{slowrollbasic}
& \epsilon_1=-\frac{\dot
H}{H^2},\,\,\,\epsilon_2=\frac{\ddot\phi}{H\dot\phi},
\,\,\,\epsilon_4=\frac{\dot E}{2HE},
\\ \notag &
\epsilon_5=\frac{Q_a}{2HQ_t},\,\,\, \epsilon_6=\frac{\dot
Q_t}{2HQ_t}\, ,
\end{align}
with $E$ is defined as follows,
\begin{equation}\label{functionE}
E=\frac{F}{\dot\phi^2}\left(
\dot{\phi}^2+3\left(\frac{Q_a^2}{2Q_t}\right)\right)\, ,
\end{equation}
where $Q_a$, for the Einstein-Gauss-Bonnet theory is
\cite{Hwang:2005hb},
\begin{align}\label{qis}
& Q_a=-4 \dot\xi H^2\, .
\end{align}
Employing the simplified equations of motion (\ref{motion5})-
(\ref{motion8}), the slow-roll indices finally become,
\begin{equation}
\label{index1} \centering \epsilon_1\simeq\frac{\kappa^2
}{2}\left(\frac{\xi'}{\xi''}\right)^2\, ,
\end{equation}
\begin{equation}
\label{index2} \centering
\epsilon_2\simeq1-\epsilon_1-\frac{\xi'\xi'''}{\xi''^2}\, ,
\end{equation}
\begin{equation}
\label{index4} \centering
\epsilon_4\simeq\frac{\xi'}{2\xi''}\frac{\mathcal{E}'}{\mathcal{E}}\,
,
\end{equation}
\begin{equation}
\label{index5} \centering
\epsilon_5\simeq-\frac{\epsilon_1}{\lambda}\, ,
\end{equation}
\begin{equation}
\label{index6} \centering \epsilon_6\simeq
\epsilon_5(1-\epsilon_1)\, ,
\end{equation}
with, $\mathcal{E}=\mathcal{E}(\phi)$ and $\lambda=\lambda(\phi)$
being defined as follows,
\begin{equation}\label{functionE}
\mathcal{E}(\phi)=\frac{1}{\kappa^2}\left(
1+72\frac{\epsilon_1^2}{\lambda^2} \right),\,\, \,
\lambda(\phi)=\frac{3}{4\xi''\kappa^2 V}\, .
\end{equation}
Regarding the observational indices, we have \cite{Hwang:2005hb},
\begin{equation}
\label{spectralindex} \centering
n_{\mathcal{S}}=1-4\epsilon_1-2\epsilon_2-2\epsilon_4\, ,
\end{equation}
for the spectral index of the primordial scalar perturbations,
while the tensor spectral index is \cite{Oikonomou:2021kql},
\begin{equation}\label{tensorspectralindexfinal}
n_{\mathcal{T}}\simeq -2\epsilon_1\left ( 1-\frac{1}{\lambda
}+\frac{\epsilon_1}{\lambda}\right)\, .
\end{equation}
Finally, the tensor-to-scalar ratio is,
\begin{equation}\label{tensortoscalarratiofinal}
r\simeq 16\epsilon_1\, .
\end{equation}
In the next section, we shall use the equations and expressions of
this section, in order to employ and introduce the bottom-up
reconstruction method for Einstein-Gauss-Bonnet theories.

\section{Reconstruction of Inflationary Phenomenology from the Observational Indices: A Bottom-up Approach}

For our general solution it is essential to express every variable
as a function of $N$, which is the number of $e$-foldings. To
achieve that we write,
\begin{equation}
\label{xiofN} \centering \xi'(\phi) = \frac{d N}{d
\phi}\frac{d\xi}{dN} ,
\end{equation}
and
\begin{equation}
\xi^{\prime \prime}=\frac{d \xi^{\prime}}{d \phi}=\frac{d N}{d
\phi} \frac{d}{d N}\left(\frac{d N}{d \phi} \frac{d \xi}{d
N}\right)=\left(\frac{d N}{d \phi}\right)^{2} \frac{d^{2} \xi}{d
N^{2}}+\xi^{\prime} \frac{d}{d N}\left(\frac{d N}{d \phi}\right)
,
\end{equation}
where the ``prime'', denotes differentiation with respect to the
scalar field. Using Eq. (\ref{index1}),
(\ref{tensortoscalarratiofinal}) and
\begin{equation}
\label{xiprimeprimeoverxiprime} \centering
\frac{\xi''}{\xi'}=\frac{d N}{d \phi}
,
\end{equation}
which follows from Eq. (\ref{xiofN}) we obtain,
\begin{equation}
\label{finaltensortoscalar} \centering r(N)=8\kappa^2
\left(\frac{d \phi}{d N}\right)^2 ,
\end{equation}
which is the general form of the tensor-to-scalar ratio as a
function of the number of $e$-foldings, in the
Einstein-Gauss-Bonnet theory. The previous expression can be
written as,
\begin{equation}
\label{Nprime} \centering
\frac{d N}{d\phi}=\frac{2 \kappa \sqrt{2}}{\sqrt{r}}
,
\end{equation}
and thus we obtain a useful expression for $\xi'$ and $\xi''$
\begin{equation}
\centering
\xi'=\frac{d N}{d \phi}\frac{d \xi}{d N}=\frac{2\kappa\sqrt{2}}{\sqrt{r}}\frac{d \xi}{d N}
,
\end{equation}
\begin{equation}
\xi^{\prime \prime}=\frac{8 \kappa^{2}}{r} \frac{d^{2} \xi}{d N^{2}}-\frac{4 \kappa^{2}}{r^{2}} \frac{d r}{d N} \frac{d \xi}{d N}
.
\end{equation}
Using the previous equations and the fact that $\xi''=\frac{d N}{d
\phi}\xi'$ we have,
\begin{equation}
\label{xiprimeprimefinal} \xi'' = \frac{8 \kappa^{2}}{r}
\frac{d^{2} \xi}{d N^{2}}-\frac{4 \kappa^{2}}{r^{2}} \frac{d r}{d
N} \frac{d \xi}{d N}=\frac{2 \kappa \sqrt{2}}{\sqrt{r}} \frac{2
\kappa \sqrt{2}}{\sqrt{r}} \frac{d \xi}{d N}=\frac{8
\kappa^{2}}{r} \frac{d \xi}{d N} .
\end{equation}
From Eq. (\ref{xiprimeprimefinal}) we derive the differential
equation of the coupling function with respect to the number of
$e$-foldings,
\begin{equation}
\label{difxi} \centering
\frac{d^2 \xi}{d N^2}-\left(\frac{1}{2r}\frac{d r}{d N}+1\right)\frac{d \xi}{d N}=0
,
\end{equation}
and its solution is,
\begin{equation}
\label{genxi} \centering
\xi(N)=C_1 \int \sqrt{r(N)}e^{N}d N+C_2
,
\end{equation}
where $C_1$, $C_2$ are two arbitrary integration constants. From
Eq. (\ref{maindiffeqnnew}), the potential of the scalar field
takes the following form,
\begin{equation}
\label{Vwithphi}
\frac{1}{V^{2}} \frac{d V}{d \phi}+\frac{4 \kappa^{4}}{3} \frac{d \xi}{d \phi}=0
.
\end{equation}
Combining Eqs. (\ref{Vwithphi}), (\ref{xiofN}) we get,
\begin{equation}
\label{VwithN}
\frac{1}{V^{2}} \frac{d N}{d \phi} \frac{d V}{d N}+\frac{4 \kappa^{4}}{3} \frac{d N}{d \phi} \frac{d \xi}{d N}=0
.
\end{equation}
Eq. (\ref{VwithN}) is a differential equation satisfied by the
potential of the scalar field with respect to the number of
$e$-foldings, $N$. Its general solution is,
\begin{equation}
\label{genpot} \centering
V(N)=\frac{3}{4\kappa^4}\frac{1}{\xi(N)}
.
\end{equation}
Thus employing this bottom-up reconstruction method we just
presented, we are able to derive the scalar coupling function, the
potential of the scalar field and the spectral indices
$n_\mathcal{S}$, $n_T$ for various expressions of the
tensor-to-scalar ratio as a function of the number of
$e$-foldings. In the next section we shall present several
illustrative examples of viable Einstein-Gauss-Bonnet inflationary
models, which are derived by employing our bottom-up
reconstruction approach.

\section{Application of the Bottom-up Formalism: Phenomenology of Various Models}

We proceed by exploring the phenomenology of different models,
using the method we developed in the previous section. For most of
our models we shall choose a tensor-to-scalar ratio of the form
$r=\delta/N^d$ where $d>0$. By varying the parameter $\delta$, so
that the tensor-to-scalar ratio is within the Planck 2018
\cite{Planck:2018jri} constraints $r < 0.056$, we solved the
equation $n_{\mathcal{S}}(C_{1})=0.9649 \pm 0.0042$ for $C_2=1$.
For each set of the parameters $\delta$, and $n_{\mathcal{S}}$
there are three different values for the parameter $C_1$, which
verify the equation and each of these gives a different $n_T$: a
negative one (about -0.04) and two positive (about 0.04 and 0.92)
ones. In all the cases for which $r=\delta/N^d$, the parameter
$C_1$ turns out to be $C_1\mathcal{O}(10^{-27}-10^{-24})$ in order
for the models to be rendered viable. The last of our models is an
exponential model of the form of $r=a e^{-b N}$. The methodology
is the same with the previous models, only here the free
parameters of the model are $a$ and $b$. In our study the value of
the integration constant $C_{2}$ did not affect any of the
calculated indices, and so we kept $C_2=1$ in every model. For
brevity we will not show the evaluation of the slow-roll indices
and the analytical expression of the scalar spectral index
$n\mathrm{S}$, since these are too lengthy.

\subsection{Model I: The Case with $r=\delta/N$}

\begin{figure}[h!]
\centering
\includegraphics[width=40pc]{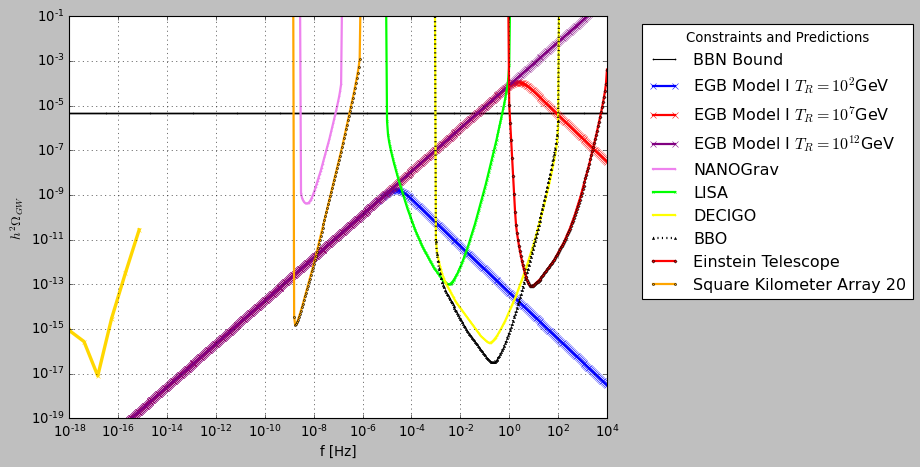}
\caption{ The $h^2$-scaled primordial gravitational waves energy
spectrum for the Model I, versus the sensitivity curves of future
primordial gravitational waves experiments. }\label{ModelI}
\end{figure}

We start with the simplest possible model of our study, in which
case,
\begin{equation}
\label{r1} \centering
r=\frac{\delta}{N}
.
\end{equation}
\begin{figure}[h!]
\centering
\includegraphics[width=20pc]{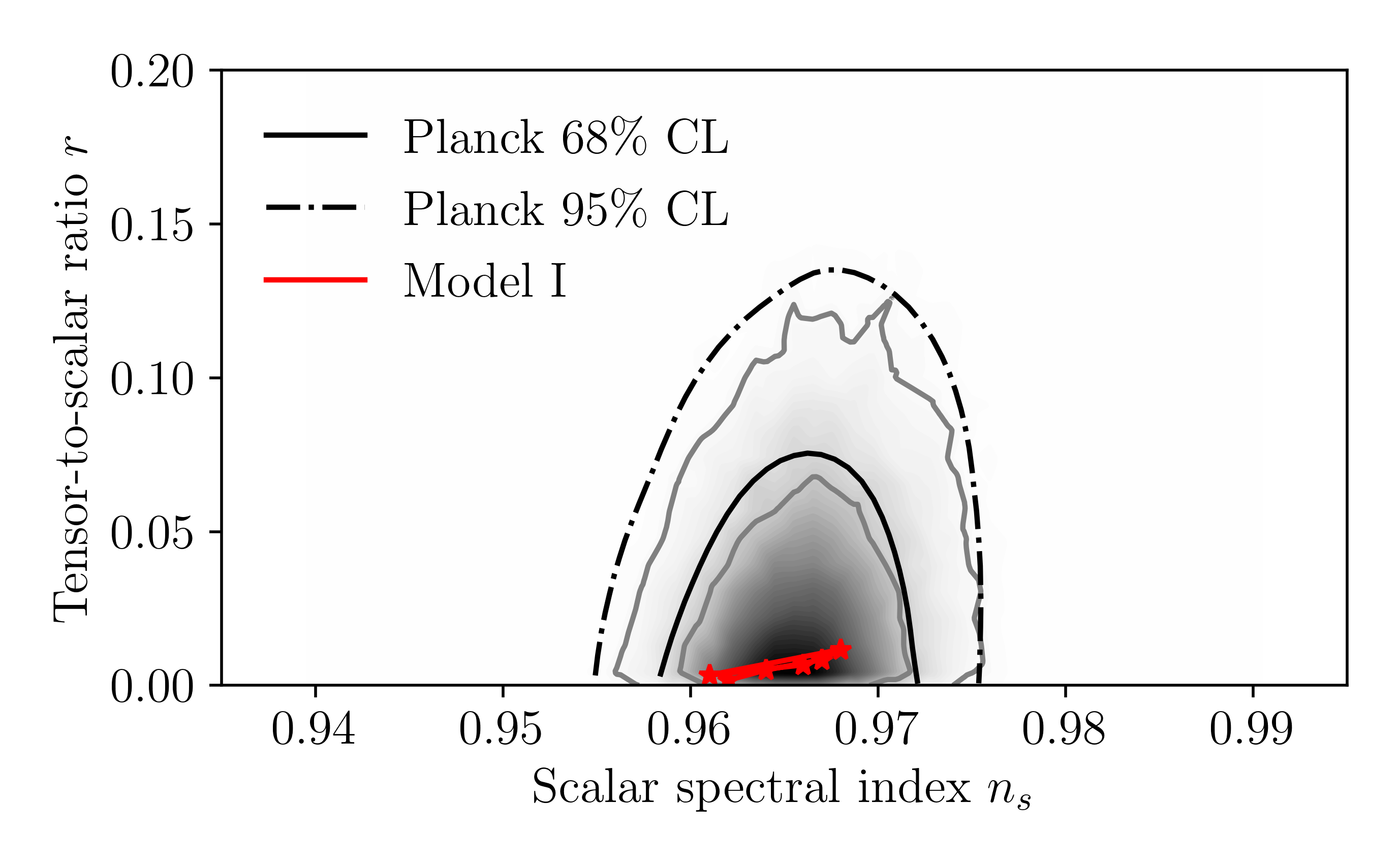}
\caption{Planck 2018 Likelihood Curves for the Model I with
$r=\delta/N$ model}\label{plotplanck2}
\end{figure}
From Eq. (\ref{genxi}) it follows that the scalar coupling
function $\xi$ takes the form,
\begin{equation}
\label{xi1} \centering
\xi=2 C_1 \sqrt{\delta} e^{N} \operatorname{D}(\sqrt{N})+C_2
,
\end{equation}
where $\operatorname{D}(x)$ is the Dawson integral, defined as
$\operatorname{D}(x)=e^{-x^2} \int_0^x e^{t^2}d t$. From
Eq.  (\ref{genpot}) it follows that the potential of the scalar
field $V$ takes the form,
\begin{equation}
\label{V1} \centering
V(N)=\frac{3}{4\kappa^4 \left(2 C_1 e^N \sqrt{\delta} \operatorname{D}\left(\sqrt{N}\right)+C_2 \right)}
,
\end{equation}
To calculate the scalar spectral index, the corresponding value of
the tensor-to-scalar ratio and the tensor spectral index, we need
to calculate the slow-roll indices. To do that we use the function
$\lambda(N)$ (\ref{functionE}), which for this model is,
\begin{equation}
\label{} \centering
\lambda(N)=\frac{C_2}{8 C_1}\sqrt{\frac{\delta}{N}}e^{-N}+\frac{\delta \operatorname{D}(\sqrt{N})}{4 \sqrt{N}}
.
\end{equation}
Using Eq. (\ref{tensorspectralindexfinal}) the tensor spectral
index takes the form,
\begin{equation}
\label{nT1} \centering n_T=-\frac{2 C_2 \delta \sqrt{N}+C_1
\sqrt{\delta} e^N \left(\delta - 16N 4 \delta \sqrt{N}
\operatorname{D}(\sqrt{N})\right)}{16 C_2 N^{\frac{3}{2}}+32 C_1
\sqrt{\delta} e^{N} N^{\frac{3}{2}}
\operatorname{D}(\sqrt{N})} .
\end{equation}
The upper limit of the $\delta$ parameter, in order for the
tensor-to-scalar ratio $r$ to comply with $2018$ Planck
constrains, is $3.36$. By varying the parameter $\delta$,
$n_{\mathcal{S}}$, $C_1$ and $C_2$ with respect to the Planck
constraints we evaluated the tensor-to-scalar ratio and the tensor
spectral index. In the Table \ref{tab:Model 1} we present a small
portion of the set of values of the observational indices we
calculated.
\begin{table}[h!]
  \begin{center}
    \caption{Different values for $\delta$ and $n_{\mathcal{S}}$ and the corresponding $n_T$ and $r$ for Model I}
\label{tab:Model 1}
    \begin{tabular}{|r|r|r|r|}
     \cline{1-4}
$\delta$    & $r$          & $n_{\mathcal{S}}$  & $n_T$      \\
\hline
0.001       & $1.66667\cdot10^{-5}$  & 0.965              & 0.964288 \\
\hline
0.05        & 0.000833333 & 0.965               & 0.964191 \\
\hline
0.1         & 0.00166667  & 0.965               & 0.039063 \\
\hline
1           & 0.016667    & 0.961               & 0.037994 \\
\hline
3           & 0.05        & 0.969               & 0.960455 \\
\hline
    \end{tabular}
  \end{center}
\end{table}
Using the data presented in the Table \ref{tab:Model 1} we
confront the model with the Planck 2018 data
\cite{Planck:2018jri}, and the results are presented in Fig.
\ref{plotplanck2}. As it can be seen in Fig. \ref{plotplanck2},
Model I is well fitted deeply in the Planck likelihood curves.
Also in Fig. \ref{ModelI}
we plot the $h^2$-scaled primordial gravitational waves energy
spectrum for the Model I, versus the sensitivity curves of future
primordial gravitational waves experiments for three reheating
temperatures. We chose the smallest value of the predicted
tensor-to-scalar ratio and the corresponding blue-tilted spectral
index. As it can be seen, the predicted energy spectrum lies
within the reach of future experiments.

\subsection{Model II: The Case with $r=\delta/N^2$}

\begin{figure}[h!]
\centering
\includegraphics[width=40pc]{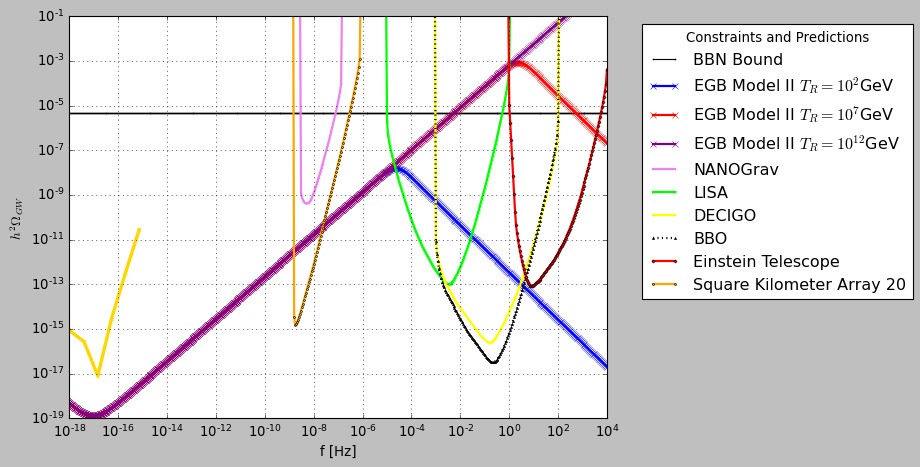}
\caption{ The $h^2$-scaled primordial gravitational waves energy
spectrum for the Model II, versus the sensitivity curves of future
primordial gravitational waves experiments. }\label{ModelII}
\end{figure}

The next model we shall consider has the following
tensor-to-scalar ratio,
\begin{equation}
\label{r2} \centering
r=\frac{\delta}{N^2}
.
\end{equation}
Again we calculate $\xi$ from Eq. (\ref{genxi}),
\begin{equation}
\label{xi2} \centering \xi=2 C_1 \sqrt{\delta}
\operatorname{Ei}(N)+C_2 ,
\end{equation}
where $\operatorname{Ei}(x)$ is the Exponential integral, defined
as $\operatorname{Ei}(x)=\int_{-\infty}^x\frac{e^{t}}{t}d t$
\begin{figure}[h!]
\centering
\includegraphics[width=20pc]{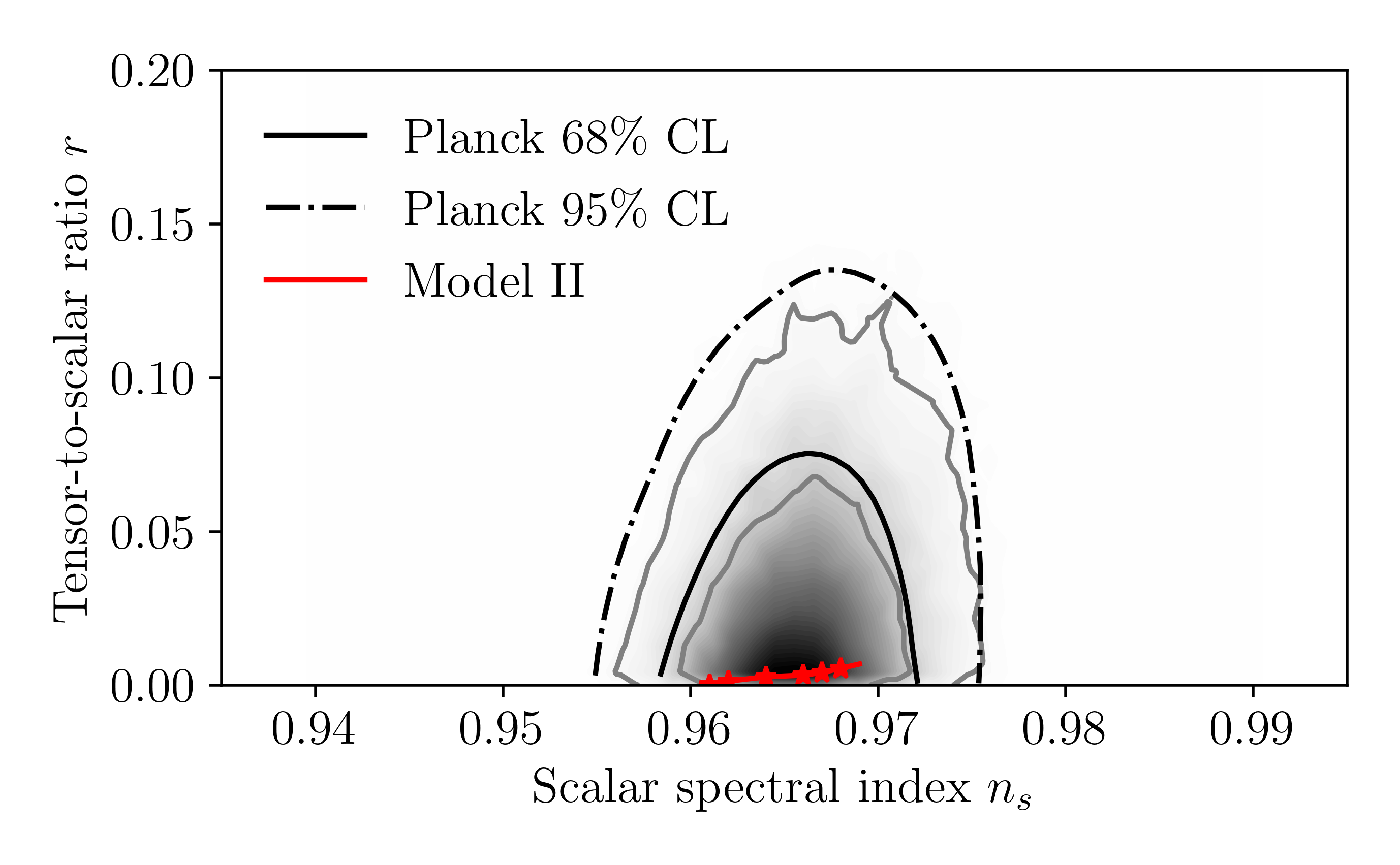}
\caption{Planck 2018 Likelihood Curves for the Model II for
$r=\delta/N^2$.}\label{plotplanck3}
\end{figure}
We proceed to the calculation of the potential of the scalar field
$V$ by applying Eq. (\ref{genpot}),
\begin{equation}
\label{V2} \centering
V(N)=\frac{3}{4\kappa^4 \left(2 C_1 \sqrt{\delta} \operatorname{Ei}(N)+C_2\right)}
,
\end{equation}
and he function $\lambda(N)$ is,
\begin{equation}
\label{} \centering
\lambda(N)=\frac{C_2 \sqrt{\delta}}{8 C_1}\frac{e^{-N}}{N}+\frac{\delta}{8}\frac{e^{-N}\operatorname{Ei}(N)}{N}
.
\end{equation}
From Eq. (\ref{tensorspectralindexfinal}) it follows that the
tensor spectral index $n_T$ takes the form,
\begin{equation}
\label{} \centering
n_T=-\frac{\sqrt{\delta} \left(2 C_2 \sqrt{\delta} N+ C_1 e^{N} \left(\delta-16 N^2\right)+2 C_1 \delta N \operatorname{Ei}(N)\right)}{16N^3 \left(C_2+C_1 \sqrt{\delta} \operatorname{Ei}(N) \right)}
.
\end{equation}
The upper limit of the $\delta$ parameter, in order for the
tensor-to-scalar ratio $r$ to comply with $2018$ Planck
constraints, is $201.6$. In Table \ref{tab:Model 2} we present
some of the values of the observational indices we calculated.
\begin{table}[h!]
  \begin{center}
    \caption{Different values for $\delta$ and $n_{\mathcal{S}}$ and the corresponding $n_T$ and $r$ for Model II}
\label{tab:Model 2}
    \begin{tabular}{|r|r|r|r|}
     \cline{1-4}
$\delta$  & $r$          & $n_{\mathcal{S}}$  & $n_T$      \\
\hline
1         & 0.000277778  & 0.961              & 0.944884\\
\hline
5         & 0.00138889   & 0.962              & 0.945287 \\
\hline
10        & 0.00277778   & 0.964              & 0.946191 \\
\hline
15        & 0.00416667   & 0.967              & 0.947627 \\
\hline
20        & 0.00555556   & 0.968              & 0.947996 \\
\hline
    \end{tabular}
  \end{center}
\end{table}
Using the data presented in Table \ref{tab:Model 2} we confront
the model with the Planck 2018 data \cite{Planck:2018jri}, and the
results are presented in Fig. \ref{plotplanck3}. As it can be seen
in Fig. \ref{plotplanck3}, Model II is well fitted deeply in the
Planck likelihood curves, as in the previous model. Also in Fig.
\ref{ModelII}
we plot the $h^2$-scaled primordial gravitational waves energy
spectrum for the Model II, versus the sensitivity curves of future
primordial gravitational waves experiments for three reheating
temperatures. In this case too we chose the smallest value of the
predicted tensor-to-scalar ratio and the corresponding blue-tilted
spectral index. As in the previous case, the predicted energy
spectrum lies within the reach of future experiments.

\subsection{Model III: The case with $r=\delta/N^3$}

\begin{figure}[h!]
\centering
\includegraphics[width=40pc]{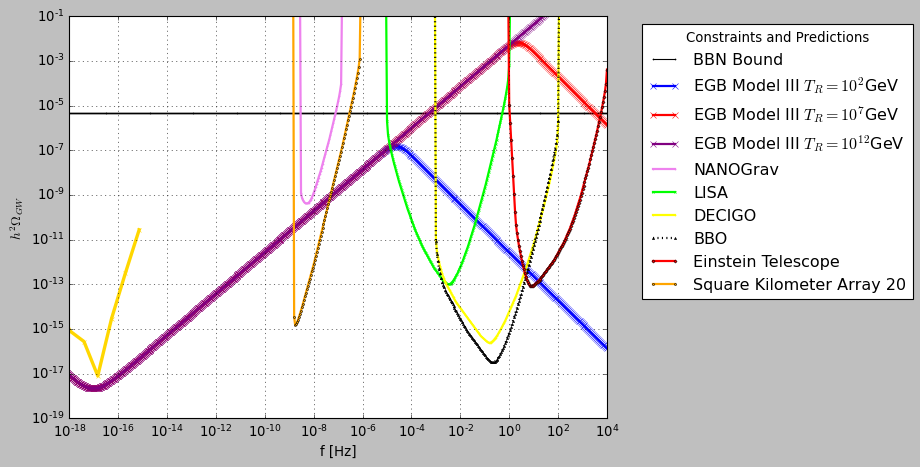}
\caption{ The $h^2$-scaled primordial gravitational waves energy
spectrum for the Model III, versus the sensitivity curves of
future primordial gravitational waves experiments.
}\label{ModelIII}
\end{figure}

In this model the expression of the tensor-to-scalar ratio is,
\begin{equation}
\label{r3} \centering
r=\frac{\delta}{N^3}
.
\end{equation}
The scalar coupling function of this model $\xi(N)$ is,
\begin{equation}
\label{xi3} \centering \xi(N) =C_{2}-2C_{1}
\sqrt{\frac{\delta}{N^{3}}} N\left(e^{N}+N
\operatorname{E}_{\frac{1}{2}}(-N)\right) ,
\end{equation}
where $\operatorname{E}_n(x)$ is the Exponential integral function, defined as $\operatorname{E}_n(x)=\int_1^{\infty}\frac{e^{-x t}}{t^n}d t$
\begin{figure}[h!]
\centering
\includegraphics[width=20pc]{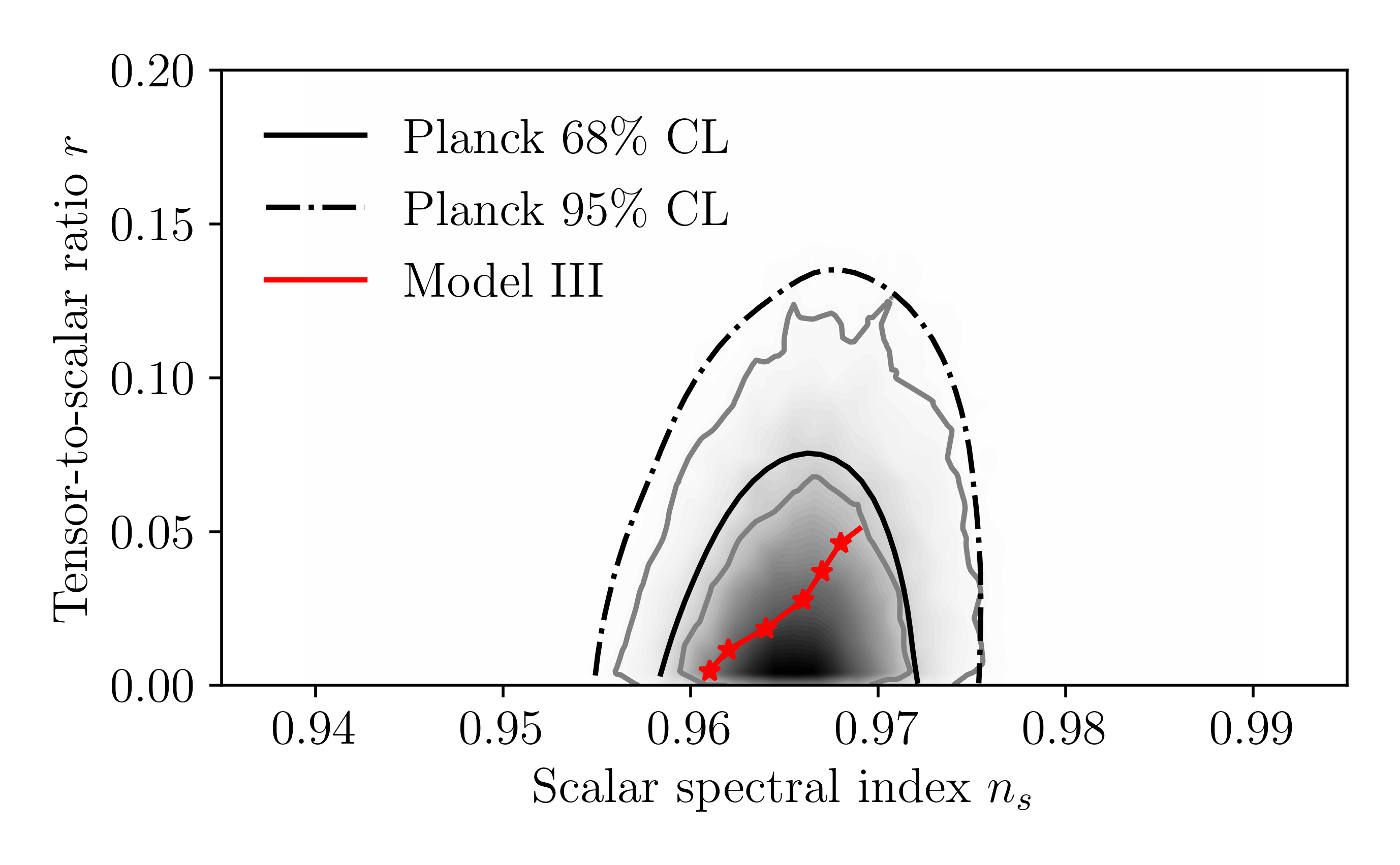}
\caption{Planck 2018 Likelihood Curves for the Model III for
$r=\delta/N^3$.}\label{plotplanck4}
\end{figure}
The potential of the scalar field is,
\begin{equation}
\centering
V(N) = \frac{3}{4 \kappa^{4}\left(C_{2}-2 C_{1} \sqrt{\frac{6}{N^{3}}} N\left(e^{N}+N \operatorname{E}_{\frac{1}{2}}\left(-N\right)\right)\right)}
,
\end{equation}
and the $\lambda(N)$ function is,
\begin{equation}
\centering
\lambda(N) = \frac{e^{-N}\left(-2 C_{1} e^{N} \delta+C_{2} \sqrt{\frac{\delta}{N^{3}}} N^{2}-2 C_{1} \delta N \operatorname{E}_{\frac{1}{2}}\left(-N\right)\right)}{8 C_{1} N^{2}}
.
\end{equation}
Once again for brevity we will present the full forms of the
slow-roll indices and the analytic expression of the
$n_\mathcal{S}$. The tensor spectral index of this model is,
\begin{equation}
\centering n_T = \frac{\delta\left(-\frac{2
C_{2}\delta}{\sqrt{\frac{\delta}{N^{3}}}}+C_{1}
e^{N}\left(-\delta+4 \delta N+16 N^{3}\right)+4 C_{1} \delta N^{2}
\operatorname{E}_{\frac{1}{2}}\left(-N\right)\right)}{16
N^{4}\left(-2 C_{1} e^{N} \delta+C_{2} \sqrt{\frac{\delta}{N^{3}}}
N^{2}-2 C_{1} \delta N
\operatorname{E}_{\frac{1}{2}}\left(-N\right)\right)}.
\end{equation}
The 2018 Planck Data constrain the scalar spectral index and thus
the value of the parameter $\delta$. In this case $\delta <
12096$. Varying the parameter $\delta$, and $n_\mathcal{S}$, we
are able to calculate the exact value of the tensor-to-scalar
ratio $r$ and the tensor spectral index $n_{T}$. Some of the
different values we calculated to make the likelihood curve of
this model are shown in Table \ref{tab:Model 3}.
\begin{table}[h!]
\caption{Different values for $\delta$ and $n_{\mathcal{S}}$ and
the corresponding $n_T$ and $r$ for Model III.} \label{tab:Model
3}
\begin{tabular}{|l|l|l|l|}
\cline{1-4}
$\delta$   & r   & $n_{\mathcal{S}}$    & $n_{\mathcal{T}}$              \\ \cline{1-4}
1000  & 0.00463  & 0.961 & 0.92709     \\ \cline{1-4}
2500  & 0.011574 & 0.962 & 0.926818    \\ \cline{1-4}
6000  & 0.027778 & 0.966 & 0.927071    \\ \cline{1-4}
8000  & 0.037037 & 0.967 & 0.926526    \\ \cline{1-4}
11000 & 0.050926 & 0.969 & 0.92597     \\ \cline{1-4}
\end{tabular}
\end{table}
Using the data presented in the Table \ref{tab:Model 3} we
confront the model with the Planck 2018 data
\cite{Planck:2018jri}, and the results are presented in Fig.
\ref{plotplanck4}. As it can be seen in Fig. \ref{plotplanck4},
also Model III is well fitted deeply in the Planck likelihood
curves. Also in Fig. \ref{ModelIII}
we plot the $h^2$-scaled primordial gravitational waves energy
spectrum for the Model III, and all the sensitivity curves of
future primordial gravitational waves experiments for three
reheating temperatures. As in all the previous cases, in this case
too we chose the smallest value of the predicted tensor-to-scalar
ratio and the corresponding blue-tilted spectral index. Thus Model
III is also detectable by future gravitational waves experiments.

\subsection{Model IV: The Case with $r=\delta/N^4$}

\begin{figure}[h!]
\centering
\includegraphics[width=40pc]{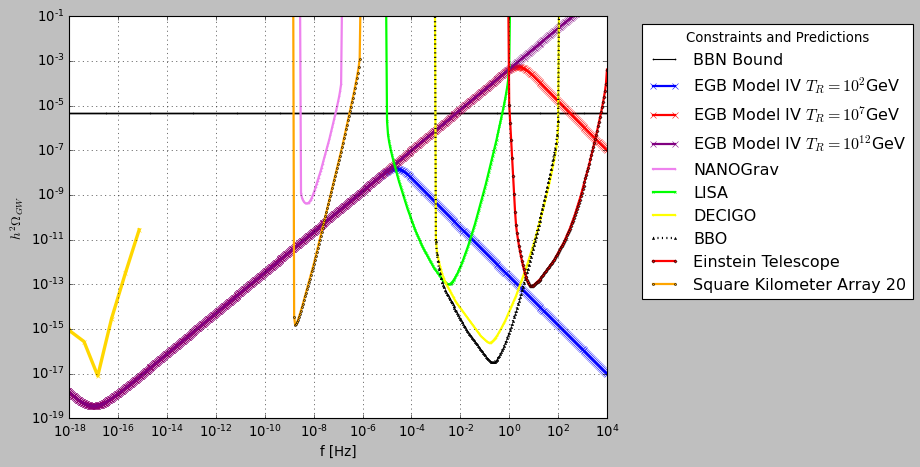}
\caption{ The $h^2$-scaled primordial gravitational waves energy
spectrum for the Model IV, versus the sensitivity curves of future
primordial gravitational waves experiments. }\label{ModelIV}
\end{figure}

Now we consider a tensor-to-scalar ratio of the form,
\begin{equation}
\centering
r=\frac{\delta}{N^4}
,
\end{equation}
for which, the scalar coupling function is,
\begin{equation}
\centering
\xi(N) =C_{2}+C_{1} \frac{\sqrt{\delta}}{N}\left(-e^{N}+N \operatorname{Ei}(N)\right)
,
\end{equation}
where $C_{2}$, $C_{1}$ are integration constants and  $N$ is the
of $e$-foldings number.
\begin{figure}[h!]
\centering
\includegraphics[width=20pc]{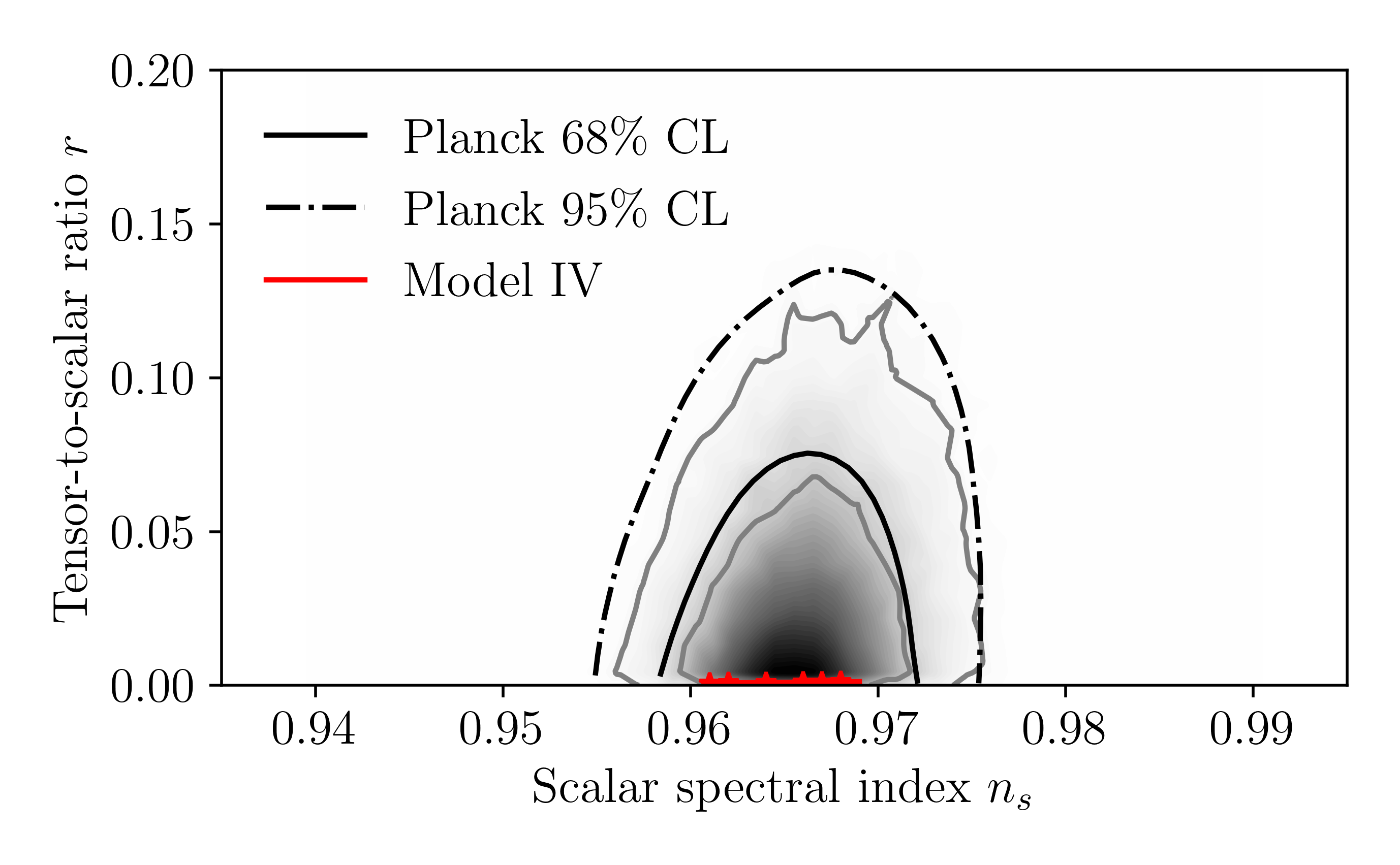}
\caption{Planck 2018 Likelihood Curves for the Model IV for
$r=\delta/N^4$.}\label{plotplanck5}
\end{figure}
The potential of the scalar field as a function of $N$ can be
found by using Eq. (\ref{genpot}) and it reads,
\begin{equation}
\centering
V(N) =\frac{3}{4 \kappa^{4}\left(C_{2}+C_{1} \sqrt{\frac{6}{N^{4}}} N\left(-e^{N}+N \operatorname{Ei}(N)\right)\right)}
.
\end{equation}
The function $\lambda$ for this model is,
\begin{equation}
\centering
\lambda(N)=\frac{e^{-N}\left(C_2 \frac{\sqrt{\delta}}{N^{2}}-\frac{C_1 e^{N} \delta}{N^{3}}+\frac{C_1 \delta \operatorname{Ei}\left(N\right)}{N^{2}}\right)}{8 C_1}
.
\end{equation}
Repeating the work we have done in the previous cases we
calculated the slow-roll indices using the above equations and
then the scalar spectral index and the tensor spectral index
(\ref{spectralindex}),(\ref{tensorspectralindexfinal}). For
brevity we show only the tensor spectral index,
\begin{equation}
\centering
n_T=\frac{\delta\left(-\frac{2 C 2 \delta}{\sqrt{\frac{6}{N^{4}}}}+C 1 e^{N}\left(-\delta+2 \delta N+16 N^{4}\right)-2 C_1 \delta N^{2} \operatorname{Ei}(N)\right)}{16 N^{5}\left(-C_1 e^{N} \delta+C 2 \sqrt{\frac{6}{N^{4}}} N^{3}+C 16 N \operatorname{Ei}(N)\right)}
,
\end{equation}
where $\delta < 725760$ in order for the spectral index to satisfy
the constraints imposed by the 2018 Planck Data. Some of the
different values of the parameter $\delta$, $r$, the spectral
index $n_\mathcal{S}$ and $n_T$ for this model are presented in
Table \ref{tab: model 4}.
\begin{table}[h!]
\caption{Different values for $\delta$ and $n_\mathcal{S}$ and the
corresponding $n_T$ and $r$ for Model IV} \label{tab: model 4}
\begin{tabular}{|l|l|l|l|}
\hline $\delta$ & $r$                                  &
$n_\mathcal{S}$    & $n_T$       \\ \hline 1000     & 0.000772 &
0.961 & 0.910203 \\ \hline 500000   & 0.0385802 & 0,966 & 0.908538
\\ \hline 100      & $7.71605 \cdot 10^{-6}$ & 0.966 & 0.912978 \\
\hline 14000    & 0.00108 & 0.967 & 0.913392 \\ \hline 16000    &
0.001223457 & 0.969 & 0.914448 \\ \hline
\end{tabular}
\end{table}
Using the data presented in the Table \ref{tab: model 4} we
confront the model with the Planck 2018 data \cite{Planck:2018jri}
in Fig. \ref{plotplanck5}. As it can be seen in Fig.
\ref{plotplanck5}, Model IV is also well fitted deeply in the
Planck likelihood curves. Also in Fig. \ref{ModelIV}
we plot the $h^2$-scaled primordial gravitational waves energy
spectrum for the Model IV, versus the sensitivity curves of future
primordial gravitational waves experiments, again for three
reheating temperatures. As in all the previous models, we chose
the smallest value of the predicted tensor-to-scalar ratio and the
corresponding blue-tilted spectral index. This model too leads to
a detectable energy power spectrum.

\subsection{Model V: The Case with $r=\delta/\sqrt{N}$}

\begin{figure}[h!]
\centering
\includegraphics[width=40pc]{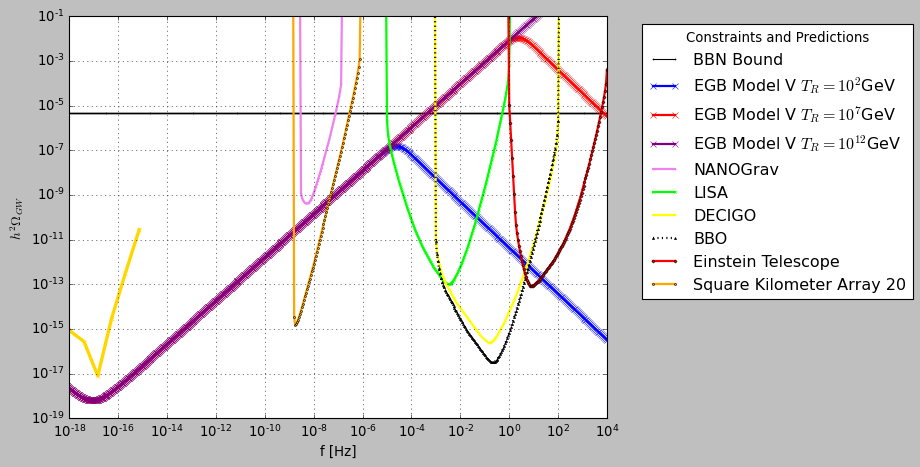}
\caption{ The $h^2$-scaled primordial gravitational waves energy
spectrum for the Model V, versus the sensitivity curves of future
primordial gravitational waves experiments. }\label{ModelV}
\end{figure}

Let us now consider non-integer powers $d$ for the
tensor-to-scalar ratio, and assume for simplicity that
$d=\frac{1}{2}$ so the tensor-to-scalar ratio has the form,
\begin{equation}
\label{} \centering
r=\frac{\delta}{\sqrt{N}}
.
\end{equation}
From Eq. (\ref{genxi}) we find,
\begin{equation}
\label{} \centering
\xi=C_2-C_1 \sqrt{\delta} N^{\frac{3}{4}} \operatorname{E}_{\frac{1}{4}}(-N)
,
\end{equation}
and the potential of the scalar field, according to Eq.
(\ref{genpot}), takes the form,
\begin{equation}
\label{} \centering
V(N)=\frac{3}{4 \kappa^4 \left(C_2-C_1 \sqrt{\delta} N^{\frac{3}{4}} \operatorname{E}_{\frac{1}{4}}(-N)\right)}
.
\end{equation}
The function $\lambda(N)$ in this case is,
\begin{equation}
\label{} \centering
\lambda(N)=\frac{C_2\sqrt{\delta}}{8 C_1}e^{-N} N^{-\frac{1}{4}}-\frac{\delta}{8}\sqrt{N}e^{-N}\operatorname{E}_{\frac{1}{4}}(-N)
.
\end{equation}
\begin{figure}[h!]
\centering
\includegraphics[width=20pc]{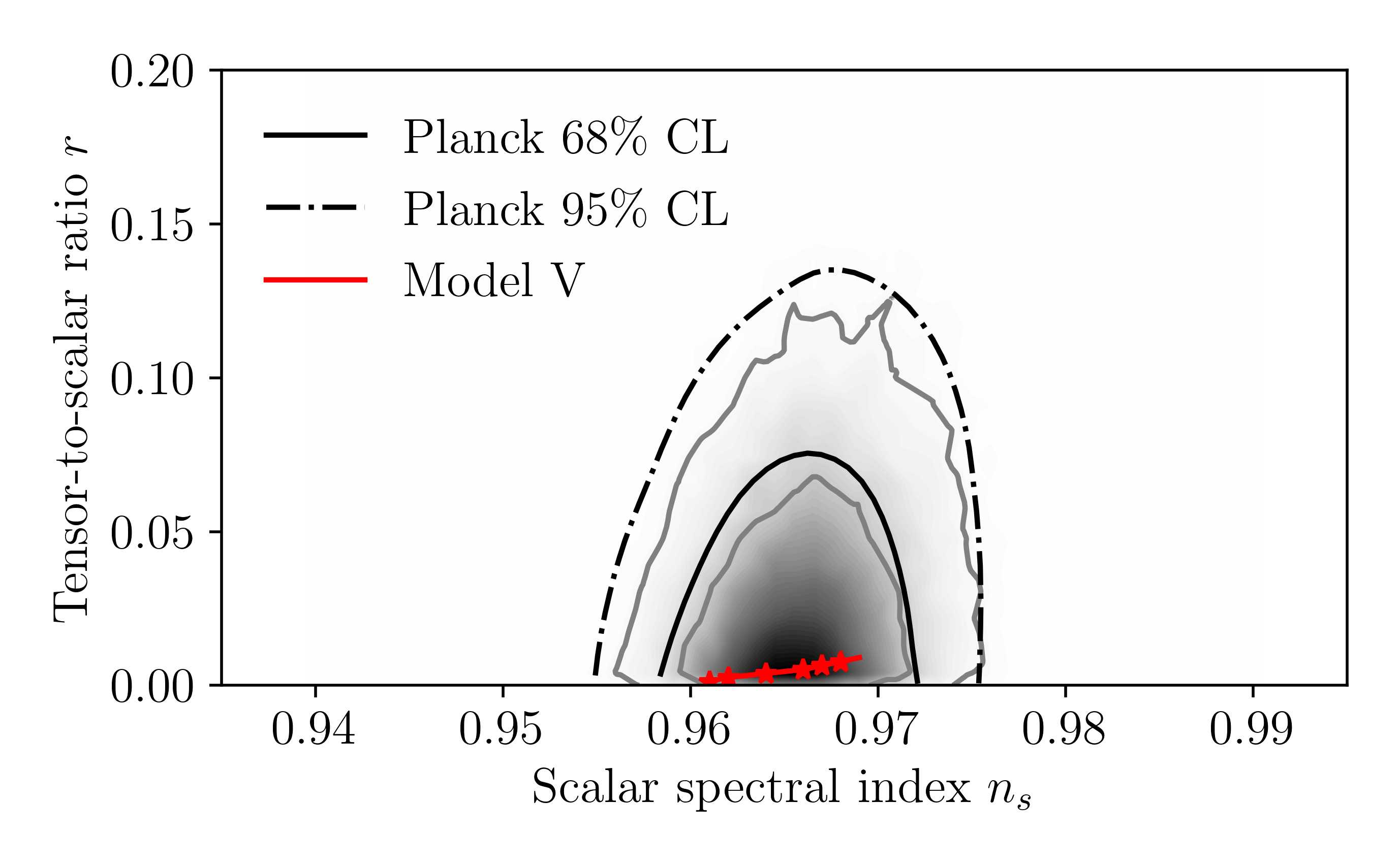}
\caption{Planck 2018 Likelihood Curves for the Model V
for$r=\delta/\sqrt{N}$.}\label{plotplanck0.5}
\end{figure}
Finally, the tensor spectral index is given by,
(\ref{tensorspectralindexfinal}),
\begin{equation}
\label{} \centering n_T=\frac{\delta \left(C_1 e^{N} \left(\delta
-16\sqrt{N}\right)+2 C_2 \right)}{16 C_1 \delta N^{\frac{3}{2}}
\operatorname{E}_{\frac{1}{4}}(-N)-16 C_2 \sqrt{\delta}
N^{\frac{3}{4}}} .
\end{equation}
The upper limit of the $\delta$ parameter, in order for the
tensor-to-scalar ratio $r$ to comply with $2018$ Planck
constrains, is $0.433774$. By varying the parameter $\delta$ and
the spectral index $n\mathcal{S}$ in Table \ref{table0.5} we
present some values of the spectral index $n\mathcal{S}$ and $n_T$
for this model.
\begin{table}[h!]
  \begin{center}
    \caption{Different values for $\delta$ and $n_{\mathcal{S}}$ and the corresponding $n_T$ and $r$ for Model V}\label{table0.5}
    \begin{tabular}{|r|r|r|r|}
     \cline{1-4}
$\delta$ & $r$         & $n_{\mathcal{S}}$  & $n_T$            \\
\hline
0.01      & 0.00129099  & 0.961              & 0.970617        \\
\hline
0.02      & 0.00258199  & 0.962              &  0.970994       \\
\hline
0.03      & 0.00387298  & 0.964              &  0.971902       \\
\hline
0.04      & 0.00516398  & 0.966              &  0.972808       \\
\hline
0.06      & 0.00774597  & 0.968              &  0.97356        \\
\hline
    \end{tabular}
  \end{center}
\end{table}
Using the data presented in the Table \ref{table0.5} we confront
the model with the Planck 2018 data \cite{Planck:2018jri} in Fig.
\ref{plotplanck0.5}. As it can be seen in Fig.
\ref{plotplanck0.5}, Model V is well fitted deeply in the Planck
likelihood curves. Also in Fig. \ref{ModelV}
we plot the $h^2$-scaled primordial gravitational waves energy
spectrum for the Model V using the smallest value of the predicted
tensor-to-scalar ratio and the corresponding blue-tilted spectral
index. As it can be seen, this model too is promising
observationally, since the predicted signal of inflationary
stochastic gravitational waves is too large.

\subsection{Exponential Model VI: The case with $r=a e^{-b N}$}

\begin{figure}[h!]
\centering
\includegraphics[width=40pc]{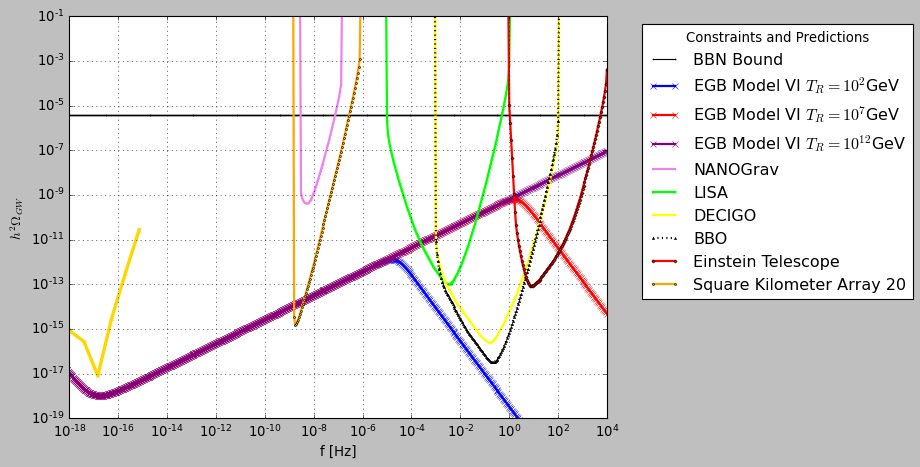}
\caption{ The $h^2$-scaled primordial gravitational waves energy
spectrum for the Model VI, versus the sensitivity curves of future
primordial gravitational waves experiments. }\label{ModelVI}
\end{figure}

The last model we studied differs from the previous ones as the
tensor-to-scalar ratio is not of the form of $r=\delta/N^d$ with
$d>0$ but of the form of $r=a e^{-b N}$, where $a$, and $b$ are
two free dimensionless parameters. Following the methodology of
the previous sections, the scalar coupling function $\xi(N)$ is,
\begin{equation}
\centering
\xi(N) = C_{2}-\frac{2 C_{1} e^{N} \sqrt{a e^{-b N}}}{b-2}.
.
\end{equation}
The potential of the scalar field as a function of the
$e$-foldings number is,
\begin{equation}
\centering
V(N) = \frac{3}{4\left(C_{2}-\frac{2 C_{1} e^{N} \sqrt{a e^{-b N}}}{-2+b}\right) \kappa^{4}},
\end{equation}
and the $\lambda(N)$ function which is essential for evaluating
the slow-roll indices of the Einstein-Gauss-Bonnet theory is,
\begin{equation}
\centering
\lambda(N) = \frac{1}{8}\left(-\frac{2 a e^{-b N}}{-2+b}+\frac{C_{2} e^{-N} \sqrt{a e^{-b N}}}{C_{1}}\right).
\end{equation}
\begin{figure}[h!]
\centering
\includegraphics[width=20pc]{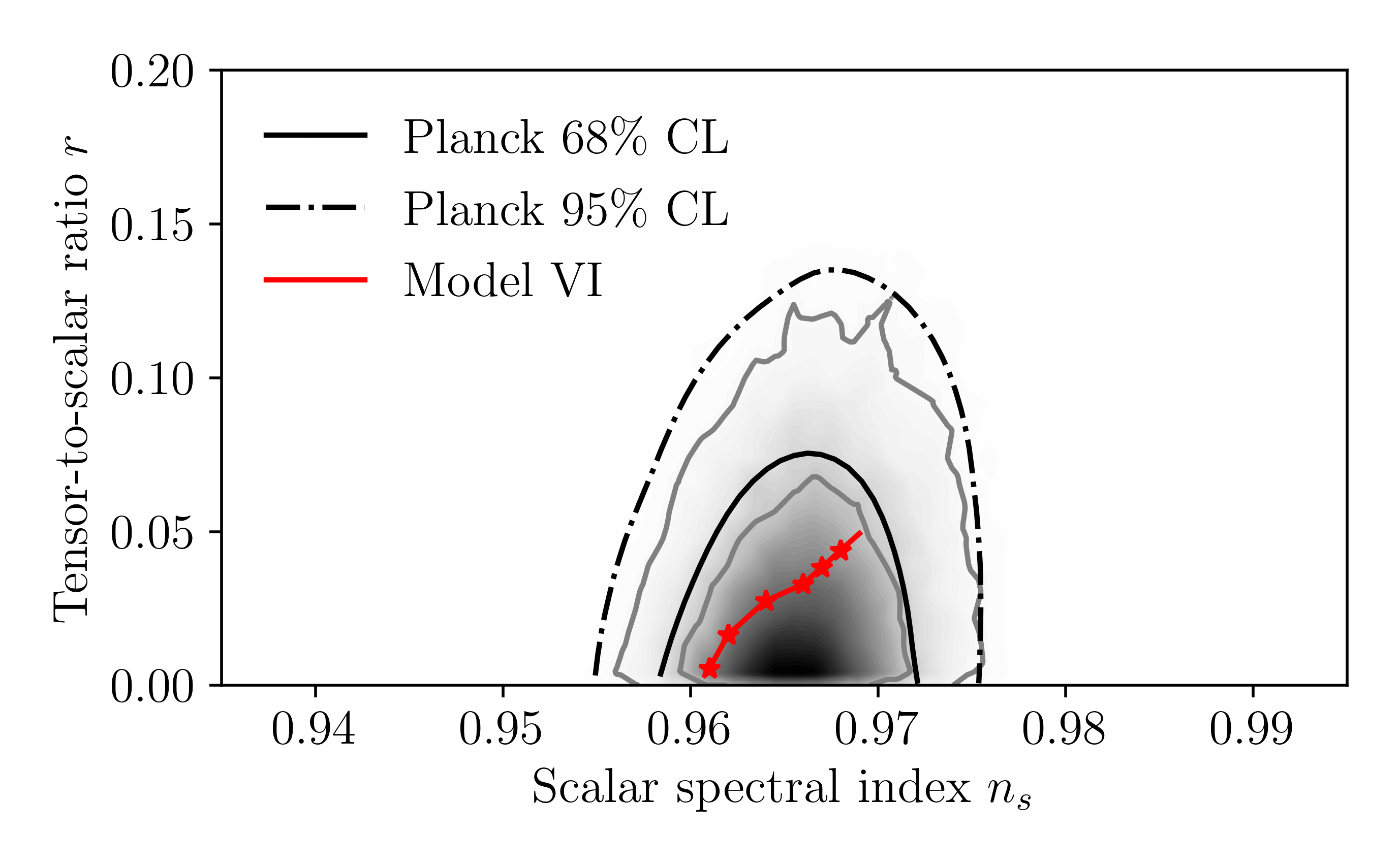}
\caption{Planck 2018 Likelihood Curves for the Model VI for $r=a
e^{-b N}$.}\label{ExpMod}
\end{figure}
Using the above equations we compute the slow roll indices and
the scalar and tensor spectral indices. The tensor spectral index
for this model is,
\begin{equation}
n_T = \frac{a\left(a(-6+b) C_{1} e^{N}-2(-2+b) e^{b N}\left(8 C_{1} e^{N}-C_{2} \sqrt{a e^{-b N}}\right)\right)}{32 a C_{1} e^{N+b N}-16(-2+b) C_{2} e^{2 b N} \sqrt{a e^{-b N}}}
\end{equation}
In our study the variation of the parameter $a$ affects the value
of the scalar to tensor ratio while the variation of the parameter
$b$ affects the value of the tensor spectral index. A set of
values of the ($n_\mathcal{S}$,$r$,$n_T$) with different values of
the parameters $a$, and $b$ are given in Table \ref{tab: exp}.
\begin{table}[h!]
\caption{Different values for $a$, $b$ and $n_\mathcal{S}$ and the
corresponding $n_T$ and $r$ for Model VI.} \label{tab: exp}
\begin{tabular}{|l|l|l|l|l|}
\hline
$a$       & $b$  & $r$        & $n_\mathcal{S}$    & $n_T$    \\ \hline
100000000 & 0.4  & 0.00377513 & 0.961              & 0.538172 \\ \hline
100000    & 0.3  & 0.001523   & 0.961                            & 0.658666 \\ \hline
100000000 & 0.43 & 0.00062402 & 0.966                            & 0.501779 \\ \hline
100000    & 0.3  & 0.001523   & 0.966                            & 0.603673 \\ \hline
100000000 & 0.4  & 0.00377513 & 0.969                            & 0.543781 \\ \hline
\end{tabular}
\end{table}
Using the data presented in the Table \ref{tab: exp} we confront
the model with the Planck 2018 data \cite{Planck:2018jri}, and the
results in this case, are presented in Fig. \ref{ExpMod}. As it
can be seen in Fig. \ref{ExpMod}, Model VI is also well fitted
deeply in the Planck likelihood curves. Also in Fig. \ref{ModelVI}
we plot the $h^2$-scaled primordial gravitational waves energy
spectrum for the Model VI, versus the sensitivity curves of future
primordial gravitational waves experiments for three reheating
temperatures. In this case too, we chose the smallest value of the
predicted tensor-to-scalar ratio and the corresponding blue-tilted
spectral index. For this exponential model, and for large
reheating temperatures, the predicted energy spectrum of the
stochastic inflationary gravitational waves can be detected by
most of the future experiments.

\section{Conclusions}

In this paper we present a bottom-up reconstruction technique for
obtaining viable inflation from GW170817-compatible
Einstein-Gauss-Bonnet theories. Based on the formalism and
approach of Ref. \cite{Oikonomou:2021kql}, the bottom-up
reconstruction technique is based on specifying the functional
form of the tensor-to-scalar ratio as a function of the
$e$-foldings number. From it we formally derived the scalar
potential and the Gauss-Bonnet scalar coupling function, as
functions of the $e$-foldings number. Accordingly, the calculation
of the spectral indices of tensor and scalar perturbations is
greatly simplified. We exemplified our new formalism by using
several functional forms for the tensor-to-scalar ratio, chosen in
such a way so that these are simple, and lead to an inflationary
phenomenology compatible with the Planck 2018 data. We thoroughly
studied several models of interest and we showed that the
compatibility with the Planck 2018 data can be achieved for a
general range of the free parameters. Furthermore, a notable
feature of most of the models is that they lead to a blue-tilted
tensor spectral index and as we showed, the energy spectrum of the
inflationary primordial gravitational waves can be detectable by
most of the future experiments on gravitational waves. Another
notable feature is that for most of the models, the
tensor-to-scalar ratio can take significantly smaller values than
most of the $R^2$-like scalar field models.

A generalization of this work should include several higher powers
of the Ricci scalar or non minimal coupling, since the presence of
the Gauss-Bonnet term compels the Jordan frame picture. The theory
would be more complicated, but it is compelling to make such
generalizations due to the fact that if string corrections are
present in the effective inflationary Lagrangian, then it is
possible that higher powers of the Ricci scalar will be present,
and specifically quadratic or cubic terms \cite{Codello:2015mba}.
With regard to non-minimal couplings, quantum corrections will be
in the form of a conformal coupling \cite{Codello:2015mba}. Thus
generalizations of this work should include conformal couplings
and possibly $R^2$ corrections separately, like in Refs.
\cite{Odintsov:2020ilr} and \cite{Odintsov:2020xji}, respectively.

\end{document}